\documentclass[apj]{emulateapj}
\usepackage{graphicx}
\usepackage{amssymb,amsmath}
\usepackage{epsfig,floatflt}

\newtheorem{theorem}{Theorem}

\newtheorem{example}[theorem]{Example}

\newtheorem{remark}[theorem]{Remark}

\newcommand{\beq}{\begin{equation}}
\newcommand{\eeq}{\end{equation}}
\newcommand{\beqar}{\begin{eqnarray}}
\newcommand{\eeqar}{\end{eqnarray}}
\newcommand{\bitemize}{\begin{itemize}}
\newcommand{\eitemize}{\end{itemize}}
\newcommand{\bsplit}{\begin{split}}
\newcommand{\esplit}{\end{split}}
\newcommand{\fnl}{f_{\mathrm{NL}}}

\shorttitle{On the linear term correction for needlets/wavelets
  NG estimators}
\shortauthors{Donzelli et al.}

\begin{document}

\title{On the linear term correction for needlets/wavelets non-Gaussianity estimators}

\author{Simona Donzelli\altaffilmark{$\star$,1}, Frode
  K. Hansen\altaffilmark{2,3}, Michele Liguori\altaffilmark{4,5},
  Domenico Marinucci\altaffilmark{6} and Sabino Matarrese\altaffilmark{4,5}}

\altaffiltext{$\star$}{donzelli@iasf-milano.inaf.it}
\altaffiltext{1}{INAF - Istituto di Astrofisica Spaziale e Fisica Cosmica Milano, Via E. Bassini 15, 20133 Milano, Italy}
\altaffiltext{2}{Institute of Theoretical Astrophysics, University of Oslo, P.O. Box 1029 Blindern, N-0315 Oslo, Norway}
\altaffiltext{3}{Centre of Mathematics for Applications, University of Oslo, P.O. Box 1053 Blindern, N-0316 Oslo, Norway}
\altaffiltext{4}{Universit\`a di Padova, Dipartimento di Fisica e Astronomia ``G. Galilei", Universit\`a degli Studi di Padova , Via Marzolo 8, 35131 Padova, Italy}
\altaffiltext{5}{INFN, Sezione di Padova, Via Marzolo 8, 35131 Padova, Italy}
\altaffiltext{6}{Dipartimento di Matematica, Universit\`a di Roma``Tor Vergata'', Via della Ricerca Scientifica 1, I-00133 Roma, Italy}

\begin{abstract}
We derive the linear correction term for needlet and wavelet
estimators of the bispectrum and the non-linearity parameter
$f_{\rm NL}$ on cosmic microwave background radiation data.
We show that on masked $WMAP$-like data with anisotropic noise, the error bars
improve by 10-20\% and almost reach the optimal error bars
obtained with the KSW estimator  \citep{ksw}.
In the limit of full-sky and isotropic noise, this term vanishes.
We apply needlet and wavelet estimators to the $WMAP$ 7-year data and obtain our best estimate $f_{\rm NL}=37.5 \pm 21.8$.
\end{abstract}


\section{Introduction}\label{sec:introduction} 

It is well known that most inflationary models predict the fluctuations in the Cosmic
Microwave Background (CMB) to be {\em close to} but {\em not exactly Gaussian}. 
Non-Gaussian predictions are strongly model dependent, thus making primordial 
non-Gaussianity (NG) a powerful tool to discriminate among different Early Universe scenarios 
(see e.g \citet{bartolorev, chenrev, liguorirev} and references therein). 

In this paper we will focus on so called {\em local} non-Gaussianity, which can be parametrized 
in the simple form:
\begin{equation}
\Phi (\mathbf{x})=\Phi _{L}(\mathbf{x})+f_{\mathrm{NL}}^{\rm local}\left( \Phi _{L}^{2}(%
\mathbf{x})-\langle \Phi _{L}^{2}(\mathbf{x})\rangle \right) \;,
\label{eqn:primordialNG}
\end{equation}%
where $\Phi (\mathbf{x})$ is the primordial curvature perturbation
field at the end of inflation and $\Phi _{L}(\mathbf{x})$ is the
Gaussian part of the perturbation. The dimensionless
parameter $f_{\mathrm{NL}}^{\rm local}$ describes the amplitude of
non-Gaussianity\footnote{For simplicity of notation in the following we will drop the superscript ``local'' and 
simply write $f_{\mathrm{NL}}$. No confusion can arise since in this context 
we are not dealing with other types of non-Gaussianity}.    
Local non-Gaussianity is predicted to arise from standard single-field slow-roll inflation
\citep{standard1,standard2}, although at a very tiny level, as well as from multi-field
inflationary scenarios, like the curvaton
\citep{moller90,curvaton1,curvaton2,curvaton3} or inhomogeneous
(pre)reheating models \citep{gamma1,gamma2,gamma3}). Even
alternatives to inflation, such as ekpyrotic and cyclic models 
\citep{ek1,ek2} predict a local NG signature. 

The expected non-Gaussian amplitude $f_{\rm{NL}}$ varies significantly from model to model. 
For example, standard single-field slow-roll inflation predicts $f_{\rm{NL}}\sim 10^{-2}$ at the end of inflation 
\citep{standard1,standard2} (and therefore a final value $\sim $
unity after general relativistic second-order perturbation effects
are taken into account \citep{second1,second2}). Such a small
value is not experimentally detectable and for this reason a 
detection of a primordial non-Gaussian signal in present and forthcoming
CMB data will rule out single-field slow-roll inflation as a
viable scenario. Motivated by these considerations many groups
have attempted to measure $f_{\rm{NL}}$ using CMB datasets,
and Wilkinson Microwave Anisotropy Probe ($WMAP$) data in particular. 

Consistently with theoretical findings, the most stringent NG bounds have been 
obtained using estimators of the CMB angular bispectrum
(namely the three-point function of CMB fluctuations in harmonic space). While in the classical 
approach to $\fnl$ estimation \citep{ksw,creminelli,amit,amit3} the starting point to build an 
optimal cubic statistic is a direct multipole expansion of the temperature field, alternative representations can be used as well, 
like e.g. the modal bispectrum expansion of \citet{FLS2009}, or bispectra of wavelet and needlet coefficients. Since all these approaches
are just based on expanding the same quantity (the angular bispectrum) in different bases (polynomial modes, wavelets, needlets etc.), 
they are also ultimately expected to yield very similar results when applied to data. This is indeed the case. For example, a recent estimate using an 
optimal bispectrum estimator has been made by 
\citet{smith}. They obtained the smallest error bars on $f_{\rm NL}$ to date, finding
$f_{\rm NL}=38 \pm 21$ on $WMAP$ 5-year data. Consistent results, although with larger error bars, were found by \citet{curto} and
\citet{pietrobon}, using parts of the bispectrum of Spherical Mexican Hat Wavelets (SMHW) \citep{martinez02} and the skewness of needlet
coefficients, and by \citet{FLS2010}, using a modal bispectrum expansion. The most updated optimal result has been found by the $WMAP$ team on the 7-year data
 with the estimate $f_{\rm NL}=32\pm 21$ \citep{komatsu11}. Wavelets provide 
again a very similar result of $f_{\rm NL}=30\pm 23$ (Fisher matrix bound $\sigma_F=22.5$) \citep{curto3}. 

At this point one could reasonably ask why it is useful to implement estimators using many different bispectrum 
representations. After all in the end they are all expected to produce basically the same output in terms of $\fnl$. 
The justification is two-fold. First of all, different expansions can provide information 
beyond $\fnl$ (like mode spectra and full bispectrum reconstruction \citep{FLS2009,FLS2010}). Second of all, 
different expansions can present important practical advantages, such as computational rapidity or 
robustness to a number of contaminants and effects (masking, non-stationarity of the noise, foreground emission 
and so on). The $WMAP$ 7-year results quoted above seem indeed to illustrate the latter point well. When we compare the 
bispectrum and the wavelet results we see that central values and error bars are both very similar. However, to achieve 
this result the bispectrum estimator needs to include a very important {\em linear correction} term. 
This additional term, originally introduced in \citet{creminelli}, subtracts from the measured three-point function 
a spurious $\fnl$ contribution due to the breaking of 
statistical isotropy introduced by masking and non-stationary noise. Without this contribution the error bars of the bispectrum 
estimator would be {\em much} larger\footnote{Note that the correction is very large for local NG, but much smaller for 
other types of NG, hence the reason to consider only local NG in this paper, which is entirely focused on issues 
related to the linear term}  than the quoted $21$ (a factor at least $4$ or $5$ larger, as shown for example in Fig.~4 
of \citet{creminelli}). It turns out however that despite having an error bar only $\sim 10 \%$ larger than the bispectrum measurement, 
the $WMAP$ 7-year 
wavelet result of $f_{\rm NL}=30.0 \pm 23$ (Fisher matrix bound $\sigma_F=22.5$) \citep{curto3} was obtained {\em without} including any linear correction. 
It seems then that the wavelet expansion is much less 
affected by masking and anisotropic noise than the bispectrum estimator is. This raises two important issues, 
firstly pointed out in \citet{FergussonShellard2011}. 
The first is why wavelets seem so much more efficient than a standard harmonic decomposition in dealing with breaking of isotropy 
in the data. This point was partly addressed in \citep{curto4}, but we think that no definite and conclusive explanation has been
provided to date (see also paragraph \ref{sec:counterexample}) . The other issue is whether it is possible to further reduce also the variance of 
wavelet-based estimators through the introduction of the linear correction term. The aim of this work is to address both these questions: firstly 
we explicitly derive the linear correction for needlet 
and wavelet estimators of the bispectrum and of the non-linearity
parameter $f_{\mathrm{NL}}$. We show how this linear term identically vanishes
for full-sky maps with isotropic noise; we also explain why this
term is in general smaller than for harmonic space based estimators, {\em although
non-negligible} (Section~\ref{theory}). We then implement the linear term correction on a needlet bispectrum estimator
and show that these results are confirmed by simulations; the 
procedures are then applied to $WMAP$ 7-year data (Section~\ref{application}).
Conclusions are drawn in Section~\ref{conclus}.

\section{Motivation for the linear term correction}\label{theory}

\subsection{Some background results}

\subsubsection{Wick products}

We recall first some well-known background facts, to fix notation. Consider
Gaussian variables $X_{1},X_{2},X_{3},$ such that $\left\langle
X_{i}\right\rangle =0,$ $\left\langle X_{i}^{2}\right\rangle =\sigma
_{i}^{2},$ $\left\langle X_{i}X_{j}\right\rangle =\sigma _{ij}.$ The Wick
product of the three variables is defined as%
\beq
:X_{1},X_{2},X_{3}:=X_{1}X_{2}X_{3}-\sigma _{12}X_{3}-\sigma _{13}X_{2}-\sigma
_{23}X_{1}\text{ .}
\eeq

\begin{example}
For $X_{1}=X_{2}=X_{3}=X$%
\[
:X,X,X:=X^{3}-3\sigma ^{2}X\text{ .}
\]%
For $\sigma ^{2}=1$ this is the well-known Hermite polynomial of order 3, $%
H_{3}(X)=X^{3}-3X.$
\end{example}

For the expected values of Wick products, the following \emph{Diagram Formula%
} holds%
\beqar \label{diagform}
&\left\langle :X_{11},X_{12},X_{13}::X_{21},X_{22},X_{23}:\right\rangle
\nonumber \\
&=\left\langle X_{11}X_{21}\right\rangle \left\langle X_{12}X_{22}\right\rangle \left\langle X_{13}X_{23}\right\rangle +5\text{
permutations }
\eeqar
where the only permutations that are considered are those such that in each
pair an element from the first triple $\left\{ X_{11},X_{12},X_{13}\right\} $
is coupled with an element from the second triple $\left\{
X_{21},X_{22},X_{23}\right\} $. It is usually convenient to visualize the
elements $\left\{ X_{11},X_{12},X_{13}\right\} $,$\left\{
X_{21},X_{22},X_{23}\right\} $ as vertices aligned on two different rows,
and the pairs as edges connecting two different vertices; the
above-mentioned diagram formula is then usually expressed by stating that
\textquotedblleft flat edges\textquotedblright\ are ruled out.

\begin{example}
For the Hermite polynomial we have
\[
\left\langle \left( H_{3}(X)\right) ^{2}\right\rangle =\left\langle \left(
X^{3}-3X\right) ^{2}\right\rangle =6\text{ .}
\]
\end{example}

We see that the expected value of the square of the third order
Wick product is much smaller than the expected value of
$\left\langle \left( X^{3}\right) ^{2}\right\rangle =15.$ In fact,
given Gaussian random variables with unit variance a standard
argument can be used to prove that Wick products yield the
smallest variance among all other polynomials of the same order.
This result is well-known and can be found in any monographs on
related subjects, see for instance \citep{mape,peta} for
two recent references; we provide here a short proof for the case
of cubic polynomials for the sake of completeness. Indeed consider
Gaussian zero mean, unit variance random variables
$X_{1},X_{2},X_{3},$ not necessarily independent, and form a
generic polynomial
\beqar
&P(X_{1},X_{2},X_{3}):=X_{1}X_{2}X_{3}+c_{1}X_{2}X_{3}
+c_{2}X_{1}X_{3}\nonumber\\
&+c_{3}X_{1}X_{2}+c_{12}X_{3}
+c_{13}X_{2}+c_{23}X_{1}+c_{123}\mbox{ ,}
\eeqar
where the $c$'s are arbitrary (fixed) real numbers. We can rewrite
\beqar
&P(X_{1},X_{2},X_{3})=:X_{1},X_{2},X_{3}:\nonumber \\
&+c_{1}X_{2}X_{3}+c_{2}X_{1}X_{3}+c_{3}X_{1}X_{2}\nonumber \\
&+(c_{12}+\left\langle X_{1}X_{2}\right\rangle)X_{3}+(c_{13}+\left\langle X_{1}X_{3}\right\rangle )X_{2}\nonumber \\
&+(c_{23}+\left\langle X_{2}X_{3}\right\rangle )X_{1}+c_{123}\mbox{,}\nonumber\\
&=:X_{1},X_{2},X_{3}:+Q(X_{1},X_{2},X_{3})\text{ ,}
\eeqar
where $Q(X_{1},X_{2},X_{3})$ is a second order polynomial in $(X_{1},X_{2},X_{3}).$ Now, it is readily seen that $:X_{1},X_{2},X_{3}:$ is
uncorrelated with any polynomial of order 1 or 2 in the same variables; for
instance
\beqar
&\left\langle :X_{1},X_{2},X_{3}:X_{1}X_{2}\right\rangle =\left\langle
X_{1}X_{2}X_{3}X_{1}X_{2}\right\rangle \nonumber\\
&-\left\langle X_{1}X_{2}\right\rangle
\left\langle X_{3}X_{1}X_{2}\right\rangle \nonumber\\
&-\left\langle X_{1}X_{3}\right\rangle \left\langle
X_{2}^{2}X_{1}\right\rangle -\left\langle X_{2}X_{3}\right\rangle
\left\langle X_{1}^{2}X_{2}\right\rangle =0\text{ ,}
\eeqar
because odd moments of Gaussian variables always vanish. Hence we have%
\beqar
&Var\left\{ P(X_{1},X_{2},X_{3})\right\}&\nonumber\\
&=Var\left\{:X_{1},X_{2},X_{3}:+Q(X_{1},X_{2},X_{3})\right\}& \nonumber\\
&=Var\left\{ :X_{1},X_{2},X_{3}:\right\} +Var\left\{
Q(X_{1},X_{2},X_{3})\right\}& \nonumber\\
&+2Cov\left\{:X_{1},X_{2},X_{3}:,Q(X_{1},X_{2},X_{3})\right\}&\nonumber\\
&=Var\left\{ :X_{1},X_{2},X_{3}:\right\} +Var\left\{
Q(X_{1},X_{2},X_{3})\right\}& \nonumber\\
&\geq Var\left\{
:X_{1},X_{2},X_{3}:\right\} \text{ ,}&
\eeqar
whence the result is established. In words, Wick polynomials of
order 3 are uncorrelated by construction with any other polynomial
of smaller order in the same random variables, and hence minimize
the variance in the class of cubic polynomials of unit coefficient
in the maximal term. This provides the heuristic rationale for the
introduction of linear correction terms in standard and
wavelet/needlet bispectrum estimators.

\subsubsection{Needlets, Mexican needlets and SMHW}\label{sec_needwav_def}

Let $b(t)$ be a weight function satisfying three conditions, namely

\begin{itemize}
\item \emph{Compact support}: $b(t)$ is strictly larger than zero only for $%
t\in \lbrack B^{-1},B],$ some $B>1$

\item \emph{Smoothness: }$b(t)$ is $C^{\infty }$

\item \emph{Partition of unity:} for all $\ell=1,2,...$ we have%
\[
\sum_{j=0}^{\infty }b^{2}\left(\frac{\ell}{B^{j}}\right)=1\mbox{ .}
\]
\end{itemize}

Recipes to construct a function $b(t)$ that satisfy these conditions are
easy to find and are provided for instance by \citep{mpbb08} and \citep{mape}.

Consider now a a grid of points $\left\{ \xi _{jk}\right\} $ on
the sphere, e.g. the
HEALPix\footnote{\tt{http://healpix.jpl.nasa.gov}} centres \citep{healpix}; the needlet system is then defined by
\beq
\psi _{jk}(x)=\sqrt{\lambda _{jk}}\sum_{\ell=B^{j-1}}^{B^{j+1}}\sum_{m=-\ell}^{\ell}b\left(\frac{\ell}{B^{j}}\right)Y_{\ell m}(x)\overline{Y}_{\ell m}(\xi _{jk})\mbox{ ,}
\eeq
with the corresponding needlet coefficients provided by%
\beqar \label{dirtra}
\beta _{jk}&=&\sqrt{\lambda _{jk}}\int_{S^{2}}f(x)\psi
_{jk}(x)dx\nonumber\\
&=&\sum_{\ell=B^{j-1}}^{B^{j+1}}\sum_{m=-\ell}^{\ell}b\left(\frac{\ell}{B^{j}}\right)a_{\ell m}Y_{\ell m}(\xi _{jk})\mbox{ .}  
\eeqar

The coefficients $\left\{ \lambda _{jk}\right\} $ are proportional to the pixel area, see for instance %
\citep{bkmpBer} for more details. 
The needlet idea has been extended by \citep{gm1,gm2} with the construction of so called Mexican needlets (see \citet{sandro} 
for numerical analysis and implementation in a  cosmological framework). Loosely speaking, the idea is to replace the compactly supported kernel $b(\ell/B^j)$ 
by a smooth function of the form
\beq\label{bmex}
b\left(\frac{\ell}{B^j}\right)=\left(\frac{\ell}{B^j}\right)^{2p}\exp\left({-\frac{\ell^2}{B^{2j}}}\right) \mbox{ ,}
\eeq
for some integer parameter $p$. 
Mexican needlets have extremely good localization properties in real space, and for $p=1$ they provide at high frequencies a good approximation to the so-called 
Spherical Mexican Hat Wavelet (SMHW) construction.
The latter is exploited for instance by \cite{martinez02,curto,curto2,curto3}, to which we refer for more discussion and definitions. In short, the SMHW coefficients at location $n$ and scale $R$ are provided by
\beq \label{smhw_w}
w(n,R)=\int_{S^2}f(x)\Psi(x,n;R)dx \mbox{ ,}
\eeq
where the wavelet filter is defined as
\beq
\Psi(x,n;R)=\frac{1}{\sqrt{2\pi}}\frac{1}{N(R)}[1+\left(\frac{y}{2}\right)^2][2-\left(\frac{y}{R}\right)^2]e^{-y^2/2R^2}\mbox{ ;}
\eeq
here, $N(R)=R\sqrt{1+R^2/2+R^4/4}$ is a normalizing constant and $y=2\tan\theta/2$ represents the distance between $x$ and $n$, 
evaluated on the stereographic projection on the tangent plane at $n$; $\theta$ is the corresponding angular distance, evaluated on the spherical surface. 

\subsection{The linear correction term for the KSW estimator}

For our arguments to follow, and to allow for a proper comparison
with the results for needlet/wavelet based
estimators below, we shall provide a brief heuristic argument to
motivate the need for a linear term correction \citep{creminelli,amit2} in the well-known
KSW bispectrum estimator for the $f_{\rm NL}$ parameter In particular, let us consider any three frequencies $%
\ell_{1},\ell_{2},\ell_{3}$ satisfying standard triangular conditions, and
consider the angle-averaged bispectrum%
\beqar
B_{\ell_{1}\ell_{2}\ell_{3}}&=&\sqrt{\frac{(2 \ell_{1}+1)(2 \ell_{2}+1)(2 \ell_{3}+1)}{4\pi}}
 \left(
\begin{tabular}{lll}
$\ell_{1}$ & $\ell_{2}$ & $\ell_{3}$ \\
$0$ & $0$ & $0$%
\end{tabular}\right) %
\nonumber\\
&\times&
\sum_{\substack{m_1 m_2 m_3}}\left(
\begin{tabular}{lll}
$\ell_{1}$ & $\ell_{2}$ & $\ell_{3}$ \\
$m_{1}$ & $m_{2}$ & $m_{3}$%
\end{tabular}%
\right) a_{\ell_1m_1}a_{\ell_2m_2}a_{\ell_3m_3}\nonumber\\
&=&\int_{S^2}\sum_{m_1m_2m_3}a_{\ell_1m_1}a_{\ell_2m_2}a_{\ell_3m_3}\nonumber\\
&\times&
Y_{\ell_1m_1}(x)Y_{\ell_2m_2}(x)Y_{\ell_3m_3}(x)dx \nonumber\\
&=&\int_{S^2}T_{\ell_1}(x)T_{\ell_2}(x)T_{\ell_3}(x)dx\text{ ,}\nonumber\\
 T_{\ell}(x)&=&\sum_m a_{\ell m}Y_{\ell m}(x)\text{ ,}
\eeqar
where we have chosen a convenient (albeit non-standard) normalization for
the bispectrum to make our argument notationally simpler - these
normalizations do not affect by any means the substance of the argument.
Now, the bispectrum should be more properly written as
\beqar
B_{\ell_{1}\ell_{2}\ell_{3}}&=&\int_{S^{2}}\left\{
:T_{\ell_{1}}(x),T_{\ell_{2}}(x),T_{\ell_{3}}(x):\right\} dx \nonumber\\
&=&\int_{S^{2}}\left\{
  T_{\ell_{1}}(x)T_{\ell_{2}}(x)T_{\ell_{3}}(x)\right\} dx\nonumber\\
&-&\int_{S^{2}}\{ \Gamma _{\ell_{1}\ell_{2}}(x)T_{\ell_{3}}(x)\nonumber\\
&+&\Gamma_{\ell_{1}\ell_{3}} T_{\ell_{2}}(x)+\Gamma _{\ell_{2}\ell_{3}}
T_{\ell_{1}}(x)\} dx\text{ ,}\nonumber\\
\Gamma_{\ell_{u}\ell_{v}}(x)&=&\left\langle T_{\ell_{u}}(x)T_{\ell_{v}}(x)\right\rangle
\text{, }u,v=1,2,3\text{ ,}
\eeqar
In the presence of full sky-maps with isotropic noise we have
that $\Gamma _{\ell_{u}\ell_{v}}(x)\equiv \Gamma _{\ell_{u}\ell_{v}},$ i.e. it would be
constant over pixels, whence for instance%
\beqar
&&\int_{S^{2}}\left\{ \Gamma _{\ell_{1}\ell_{2}}(x)T_{\ell_{3}}(x)\right\} dx
=\int_{S^{2}}\left\{ \Gamma _{\ell_{1}\ell_{2}}T_{\ell_{3}}(x)\right\} dx\nonumber\\
&&=\Gamma _{\ell_{1}\ell_{2}}\int_{S^{2}}\left\{ T_{\ell_{3}}(x)\right\}dx=0\text{ ,}
\eeqar
because
\beq
\int_{S^{2}}\sum_{m}a_{\ell m}Y_{\ell m}(x)dx=\sum_{m}a_{\ell m}\int_{S^{2}}Y_{\ell m}(x)dx%
\equiv 0\text{ .}
\eeq
(assuming the monopole is zero).
On the other hand, in the presence of anisotropic noise and/or
masked maps the previous argument cannot hold, whence the linear
term does not cancel and the variance of the Wick product
(including the Wick product) is systematically smaller than any
other cubic statistic; for instance, as before
\beqar
&&Var\left\{ \int_{S^{2}\backslash M}T_{\ell}^{3}(x)dx\right\}\nonumber \\
&=&6\int_{(S^{2}\backslash M)\times (S^{2}\backslash M)}\left\langle
T_{\ell}(x)T_{\ell}(y)\right\rangle ^{3}dxdy \ \\
&+&9\int_{(S^{2}\backslash M)\times (S^{2}\backslash M)}\left\langle
T_{\ell}^{2}(x)\right\rangle \left\langle T_{\ell}^{2}(y)\right\rangle
\left\langle T_{\ell}(x)T_{\ell}(y)\right\rangle dxdy\text{ .} \nonumber
\eeqar
Here, we have used $S^{2}\backslash M$ to denote integration over the sphere $S$ minus the masked region $M$. 
In the full-sky case $M=\emptyset ,$ with isotropic noise, we have%
\beq
\left\langle T_{\ell}^{2}(x)\right\rangle =\left\{ \frac{2 \ell+1}{4\pi }\right\}
\left\{ C_{\ell}^{CMB}+C_{\ell}^{noise}\right\}
\eeq
whence the previous expression becomes
\beqar
&=&6\left\{ \frac{2 \ell+1}{4\pi }\right\} ^{3}\left\{
C_{\ell}^{CMB}+C_{\ell}^{noise}\right\} ^{3}\nonumber\\
&\times&\int_{S^{2}\times
S^{2}}P_{\ell}^{3}(x \cdot y)dxdy\nonumber\\
&+&9\left\{ \frac{2 \ell+1}{4\pi }\right\} ^{2}\left\{
C_{\ell}^{CMB}+C_{\ell}^{noise}\right\} ^{2}\nonumber\\
&\times&\int_{(S^{2}\backslash
M)\times (S^{2}\backslash M)}P_{\ell}(x \cdot y
)dxdy\nonumber\\
&=&6\left\{ \frac{2 \ell+1}{4\pi }\right\} ^{3}\left\{
C_{\ell}^{\rm CMB}+C_{\ell}^{noise}\right\} ^{3}\left(
\begin{array}{ccc}
\ell & \ell & \ell \\
0 & 0 & 0%
\end{array}%
\right) ^{2}.
\eeqar
In the previous computations, we have written $P_\ell$ for Legendre polynomials, $x \cdot y$ for scalar products, we have used the 
well-known Wigner's 3j symbol arising from the (Gaunt) integral of the third power of $P_\ell$, and we exploited the well-known fact that%
\beq
\int_{S^{2}\times S^{2}}P_{\ell}(x \cdot y)dxdy=0\text{ ,}
\eeq
whence the so-called flat edges terms vanish. This is not so in general,
though. Note, however, that%
\beqar
&&Var\left\{ \int_{S^{2}\backslash M}\left[
T_{\ell}^{3}(x)-3\left\langle T_{\ell}^{2}(x)\right\rangle
T_{\ell}(x)\right] dx\right\} \nonumber\\
&&=6\int_{S^{2}\backslash
M\times S^{2}\backslash M}\left\langle T_{\ell}(x)T_{\ell}(y)\right\rangle ^{3}dxdy%
\text{ ,}
\eeqar
so that the flat terms are cancelled, even in the presence of anisotropic
noise or masked regions.

\subsection{Needlets/wavelets Non-Gaussianity estimators}

The situation for wavelet or needlet/like
non-Gaussianity estimators is to some extent analogous to the one for the KSW
procedure. For instance, in \citep{curto2}, Eqs. (14-16), the
variance of the following statistic is considered
\beq
\frac{1}{4\pi }\frac{1}{\sigma _{i}\sigma _{j}\sigma _{k}}
\times
\int_{S^{2}}w(R_{i},n_{1})w(R_{j},n_{1})w(R_{k},n_{1})dn_{1}\text{,}
\eeq
where each $w(R_{i},n_{1})$ is the (random) SMHW coefficient at scale $%
R_{i}\ $\ and location $n_{1}$ (Eq.~(\ref{smhw_w})); this variance can clearly be written as%
\beqar
&&\frac{1}{(4\pi )^{2}}\frac{1}{\sigma _{i}\sigma _{j}\sigma _{k}}\frac{1}{%
\sigma _{r}\sigma _{s}\sigma _{t}}\nonumber\\
&\times& \int_{S^{2}\times S^{2}} \langle
w(R_{i},n_{1})w(R_{j},n_{1})w(R_{k},n_{1})\nonumber\\
&&w(R_{r},n_{2})w(R_{s},n_{2})w(R_{t},n_{2})  \rangle
dn_{1}dn_{2}
\eeqar
For full-sky maps with isotropic noise, we would have that
\beqar
&\left\langle
w(R_{i},n_{1})w(R_{j},n_{1})w(R_{k},n_{1})w(R_{r},n_{2})w(R_{s},n_{2})w(R_{t},n_{2})\right\rangle\nonumber\\
&=\left\langle w(R_{i},n_{1})w(R_{r},n_{2})\right\rangle
\left\langle w(R_{j},n_{1})w(R_{s},n_{2})\right\rangle\nonumber\\
&\times 
\left\langle w(R_{k},n_{1})w(R_{t},n_{2})\right\rangle
+5\text{ permutations,}
\eeqar
e.g., the only terms that non-vanish are those where $%
n_{1}$ and $n_{2}$ appear in the same pair. However, as discussed above, in the presence of masked regions and/or anisotropic noise the
terms with $n_{1},n_{1}$ or $n_{2},n_{2}$ do not vanish, and we should
rather write%
\beqar
&\left\langle
w(R_{i},n_{1})w(R_{j},n_{1})w(R_{k},n_{1})w(R_{r},n_{2})w(R_{s},n_{2})w(R_{t},n_{2})\right\rangle\nonumber\\
&=\left\langle w(R_{i},n_{1})w(R_{r},n_{2})\right\rangle \left\langle
w(R_{j},n_{1})w(R_{s},n_{2})\right\rangle\nonumber\\
&\times 
\left\langle w(R_{k},n_{1})w(R_{t},n_{2})\right\rangle
+14\text{ permutations,}
\eeqar
i.e., there are 9 further ``flat'' permutations (those where two terms from
the same row are coupled - there are three different way for each row to do
this). These missing terms give to the integral a contribution of the form%
\beqar
&\frac{1}{(4\pi )^{2}}\frac{1}{\sigma _{i}\sigma _{j}\sigma _{k}}\frac{1}{%
\sigma _{r}\sigma _{s}\sigma _{t}}
\times \int_{S^{2}\times S^{2}}\left\langle
w(R_{i},n_{1})w(R_{r},n_{1})\right\rangle& \nonumber\\
&\times
\left\langle w(R_{j},n_{2})w(R_{s},n_{2})\right\rangle \left\langle w(R_{k},n_{1}) w(R_{t},n_{2})\right\rangle dn_{1}dn_{2}&\nonumber\\
&=\frac{1}{(4\pi )^{2}}\frac{1}{\sigma _{i}\sigma _{j}\sigma _{k}}\frac{1}{%
\sigma _{r}\sigma _{s}\sigma _{t}}
\int_{S^{2}\times S^{2}}\left\langle
w(R_{i},n_{1})w(R_{r},n_{1})\right\rangle& \nonumber\\
&\times \left\langle
w(R_{j},n_{2})w(R_{s},n_{2})\right\rangle
\left\langle w(R_{k},n_{1}) \right. &\nonumber\\
&\times \left. w(R_{t},n_{2})\right\rangle
dn_{1}dn_{2}&
\eeqar
Now, for isotropic noise, $\left\langle w(R_{i},n_{1})w(R_{r},n_{1})\right\rangle \simeq const$,
whence the previous quantity is approximately proportional to%
\beqar
&\int_{S^{2}\times S^{2}}\left\langle
w(R_{k},n_{1})w(R_{t},n_{2})\right\rangle dn_{1}dn_{2}\nonumber\\
&=\left\langle \int_{S^{2}}w(R_{k},n_{1})dn_{1}\times
\int_{S^{2}}w(R_{t},n_{2})dn_{2}\right\rangle =0\text{ ,}
\eeqar
because%
\beq
\int_{S^{2}}w(R_{k},n_{1})dn_{1}=0
\eeq
in the absence of masked regions. So when the
celestial sphere is fully observed, it is equivalent to consider
or not the extra, ``flat'' terms with the same indexes
$n_{1},n_{2}$ in the pairs.

As stated earlier, and as for the KSW estimators, these terms are no longer identically
zero in the presence of masked regions or anisotropic noise. A linear
correction term can therefore be needed; we discuss its derivation in the
subsection below.

\subsubsection{The linear correction term}\label{sect:linterm}

According to our previous argument, it is straightforward to see how, to
decrease the variance of the wavelet cubic statistic, it is enough to change
the cubic statistic from
\beq
\int_{S^{2}}w(R_{i},n)w(R_{j},n)w(R_{k},n)dn
\eeq
to%
\beq
\int_{S^{2}}\left( :w(R_{i},n),w(R_{j},n),w(R_{k},n):\right) dn
\eeq
i.e. subtract a linear term of the form
\beqar
&&\int_{S^{2}}w(R_{i},n)w(R_{j},n)w(R_{k},n)dn \nonumber\\
&-&\int_{S^{2}}\left\langle w(R_{i},n)w(R_{j},n)\right\rangle
w(R_{k},n)dn \nonumber\\
&-&\int_{S^{2}}\left\langle w(R_{i},n)w(R_{k},n)\right\rangle
w(R_{j},n)dn \nonumber\\
&-&\int_{S^{2}}\left\langle
w(R_{k},n)w(R_{j},n)\right\rangle w(R_{i},n)dn\text{ .}
\eeqar
A similar situation exists for the needlets bispectrum \citep{needbisp,rudjord09}, which we can implement as%
\beqar\label{need_linterm}
&&\frac{1}{\sigma _{j_{1}}\sigma _{j_{2}}\sigma _{j_{3}}}
\Big\{\sum_{k} \beta _{j_{1}k}\beta _{j_{2}k}\beta _{j_{3}k}
-\sum_{k}\Gamma _{j_{1}j_{2}}(k)\beta _{j_{3}k} \nonumber\\
&-&\sum_{k}\Gamma
_{j_{1}j_{3}}(k)\beta _{j_{2}k} -\sum_{k}\Gamma
_{j_{2}j_{3}}(k)\beta _{j_{1}k_{1}} \Big\} \text{ ,}
\eeqar
where $\beta_{jk}$ are the usual coefficients for (standard or
Mexican) needlets and%
\beq
\Gamma _{j_{1}j_{2}}(k)=\left\langle \beta _{j_{1}k}\beta
_{j_{2}k}\right\rangle \text{ .}
\eeq
Of course, under isotropic noise
\beq
\Gamma _{j_{1}j_{2}}(k)=\Gamma _{j_{1}j_{2}}=\sum_{\ell}b\left(\frac{\ell}{B^{j_{1}}}\right)b\left(%
\frac{\ell}{B^{j_{2}}}\right)\frac{2\ell+1}{4\pi }C_{\ell}\text{ .}
\eeq
Note
that the linear terms have expected value zero always%
\beq \label{expvalue_linterm}
\left\langle \Gamma _{j_{1}j_{2}}(k)\beta _{j_{3}k}\right\rangle =\Gamma
_{j_{1}j_{2}}(k)\left\langle \beta _{j_{3}k_{3}}\right\rangle =0\text{ ;}
\eeq
however the observed value of these terms over one realization of the sky is
exactly equal to zero only if the sum (or the integral) is taken over the
whole sphere and the noise is isotropic (assuming again zero monopole), i.e. for SMHW
\beqar
&&\int_{S^{2}}\left\langle w(R_{i},n)w(R_{j},n)\right\rangle
w(R_{k},n)dn\nonumber\\
&=&\left\langle w(R_{i},n)w(R_{j},n)\right\rangle
\int_{S^{2}}w(R_{k},n)dn=0
\text{ , }
\eeqar
and correspondigly for the needlets
\beq
\sum_{k_{3}}\Gamma _{j_{1}j_{2}}\beta _{j_{3}k_{3}} =\Gamma
_{j_{1}j_{2}}\sum_{k_{1}}\beta _{j_{1}k_{1}}=0\text{ ,}
\eeq
i.e. when there are no masked regions and the noise is isotropic. These are the assumptions under which the behavior of the needlet bispectrum was investigated 
by \citep{needbisp}, where it was firstly introduced in the statistical literature.\\
It should
be noted, moreover, that in practical situations the contribution of the
linear term for wavelet/needlet-like bispectrum estimators will be typically
smaller than for KSW. This can be explained as follows: consider
\beqar
&&\sum_{k}\Gamma _{j_{1}j_{2}}(k)\beta _{j_{3}k}
=\sum_{k}\Gamma
_{j_{1}j_{2}}(k)\sum_{\ell m}b\left(\frac{\ell}{B^{j_{3}}}\right)a_{\ell
  m}Y_{\ell m}(\xi
_{j_{3}k_{3}}) \nonumber\\
&=&\sum_{k}\Gamma _{j_{1}j_{2}}(k)\sum_{\ell m}b\left(\frac{\ell }{B^{j_{3}}}%
\right)\int_{S^{2}}T(x)\overline{Y}_{\ell m}(x)Y_{\ell m}(\xi _{j_{3}k})dx \nonumber\\
&\simeq& \int_{S^{2}}\Gamma _{j_{1}j_{2}}(y)\sum_{\ell}b\left(\frac{\ell}{B^{j_{3}}}%
\right)T(x)P_{\ell}(x \cdot y )dxdy\text{ .}
\eeqar
Now it is a consequence of needlet concentration in pixel space that%
\beqar
&&\int_{S^{2}}\Gamma _{j_{1}j_{2}}(y)\sum_{l}b\left(\frac{\ell}{B^{j_{3}}}%
\right)\int_{S^{2}}T(x)P_{\ell}(x \cdot y )dxdy \nonumber\\
&=&\int_{S^{2}}T(x)\left[ \int_{S^{2}}\Gamma _{j_{1}j_{2}}(y)\sum_{\ell}b\left(\frac{%
\ell}{B^{j_{3}}}\right)P_{\ell}(x \cdot y )dy\right] dx \nonumber\\
&\simeq& \int_{S^{2}}T(x)\left[ \int_{N_{\varepsilon }(x)}\Gamma
_{j_{1}j_{2}}(y)\sum_{\ell}b\left(\frac{\ell}{B^{j_{3}}}\right)P_{\ell}(x \cdot y )dy\right] dx\nonumber\\
&
\eeqar
where $N_{\varepsilon }(x)$ is a small neighborhood of $x;$ now assuming
that noise is approximately constant over $B_{\varepsilon }(x),$ we can write%
\beqar
&&\int_{B_{\varepsilon }(x)}\Gamma _{j_{1}j_{2}}(y)\sum_{l}b\left(\frac{\ell}{%
B^{j_{3}}}\right)P_{\ell}(x \cdot y )dy \nonumber\\
&\simeq& \Gamma _{j_{1}j_{2}}(y)\left[ \int_{B_{\varepsilon }(x)}\sum_{\ell}b\left(%
\frac{\ell}{B^{j_{3}}}\right)P_{\ell}(x \cdot y )dy\right] \simeq 0%
\text{ ,}
\eeqar
because by localization%
\beqar
&&\int_{B_{\varepsilon }(x)}\sum_{\ell}b\left(\frac{\ell}{B^{j_{3}}}\right)P_{\ell}(x \cdot y )dy \nonumber\\
&\simeq& \int_{S^{2}}\sum_{\ell}b\left(\frac{\ell}{B^{j_{3}}}%
\right)P_{\ell}(x \cdot y )dy\simeq 0\text{ .}
\eeqar

\subsubsection{The relationship with mean subtraction}

We shall now show how, in the presence of nearly isotropic noise, the
behaviour of the linear term is well-approximated by subtracting
scale-by-scale the sky average of wavelets or needlets coefficients. Indeed,
define%
\beq
\overline{\beta }_{j}=\frac{1}{N}\sum_{k}\beta _{jk}
\eeq
then
\beqar
&&\frac{1}{N}\sum_{k}(\beta _{j_{1}k}-\overline{\beta }_{j_{1}})(\beta
_{j_{2}k}-\overline{\beta }_{j_{2}})(\beta _{j_{3}k}-\overline{\beta }
_{j_{3}})\nonumber\\
&=&\frac{1}{N}\sum_{k}\beta _{j_{1}k}\beta _{j_{2}k}\beta _{j_{3}k}-%
\overline{\beta }_{j_{1}}\left\{ \frac{1}{N}\sum_{k}\beta _{j_{2}k}\beta
_{j_{3}k}\right\}  \nonumber \\
&-&\overline{\beta }_{j_{2}}\left\{ \frac{1}{N}\sum_{k}\beta _{j_{1}k}\beta
_{j_{3}k}\right\}
 -\overline{\beta }_{j_{3}}\left\{ \frac{1}{N}\sum_{k}\beta
_{j_{1}k}\beta _{j_{2}k}\right\} \nonumber\\
&+&2\overline{\beta }_{j_{1}}\overline{\beta }_{j_{2}}\overline{\beta }%
_{j_{3}}\text{ .}
\eeqar
Now for nearly isotropic noise, e.g., if the covariance among coefficients
\beq
\Gamma_{j_{1}j_{2}}(k)=\left\langle \beta _{j_{1}k}\beta _{j_{2}k}\right\rangle \simeq \Gamma_{j_{1}j_{2}}\eeq
is nearly constant over the sky, then, for high enough $j$, it is natural to expect that
the cross-correlation among wavelet coefficients evaluated on a sky
realization will be close to the ensemble average,
\beq
\Gamma _{j_{1}j_{2}}\simeq \left\{ \frac{1}{N}\sum_{k}\beta _{j_{1}k}\beta
_{j_{2}k}\right\} \mbox{ .}
\eeq
Moreover we also expect%
\beq
\overline{\beta }_{j_{1}}\overline{\beta }_{j_{2}}\overline{\beta }%
_{j_{3}}\simeq 0\text{ ,}
\eeq
whence
\beq
\overline{\beta }_{j_{1}}\left\{ \frac{1}{N}\sum_{k}\beta _{j_{1}k}\beta
_{j_{2}k}\right\} \simeq \overline{\beta }_{j_{1}}\Gamma _{j_{2}j_{3}}\simeq
\Gamma _{j_{2}j_{3}}\frac{1}{N}\sum_{k}\beta _{jk}\text{ ,}
\eeq
and similarly for the permutation terms, whence the linear term will
be well-approximated by mean subtraction. At smaller frequencies $j$, this argument will not work,
but at these scales noise is likely to be negligible. In general, it should be noted that the
approximation will work better in cases where noise anisotropy is not
extremely relevant, and less well otherwise. Also, for higher order
polyspectra or alternative statistics (such as KSW) the equivalence between
mean subtraction and linear term correction will no longer hold.

\subsubsection{\protect\bigskip A toy counterexample}\label{sec:counterexample}

In \citet{curto4}, Section 3.2 it is claimed that the linear term is of
order $f_{\rm NL}^{3}$ and hence negligible. This is motivated on the basis of
the asymptotic uncorrelation of the wavelet coefficients, implying the validity of the
Central Limit Theorem (CLT) for the cubic (bispectrum statistic).

We do agree on the uncorrelation of the coefficients and the Central Limit
Theorem taking place; see for instance \citep{bkmpAoS}, \citep{needbisp,lan2} 
for analytic arguments in the needlets and Mexican needlets
case. We fail to see, however, why this should necessarily imply that the resulting
linear term should be negligible.

As a toy counterexample, consider a sequence of independent Gaussian random
variables $X_{i},$ with zero mean $\left\langle X_{i}\right\rangle $ and
nonconstant (e.g. anisotropic) variance $\left\langle X_{i}^{2}\right\rangle
=\sigma _{i}^{2};$ to fix ideas, we shall take $\sigma _{i}^{2}=1$ for $%
i=1,3,5,$ (i.e. odd)  $\sigma _{i}^{2}=3$ for $i=2,4,...,$ (i.e. even). With
these assumptions, we try to mimic the behavior of nearly independent
wavelet coefficients with anisotropic noise; for simplicity, we neglect the
effect of masked regions, but the argument would be analogous. Now consider
the statistic%
\begin{eqnarray*}
B_{1n} &=&\frac{1}{\sqrt{n}}\sum_{i=1}^{n}X_{i}^{3}\mbox{ , } \\
B_{2n} &=&\frac{1}{\sqrt{n}}\sum_{i=1}^{n}\left( X_{i}-\overline{X}%
_{n}\right) ^{3}\mbox{ , }\overline{X}_{n}=\frac{1}{n}\sum_{i=1}^{n}X_{i}%
\mbox{ ,} \\
B_{3n} &=&\frac{1}{\sqrt{n}}\sum_{i=1}^{n}\left( :X_{i},X_{i},X_{i}:\right) =%
\frac{1}{\sqrt{n}}\sum_{i=1}^{n}\left( X_{i}^{3}-3\sigma
_{i}^{2}X_{i}\right) \mbox{ ,}
\end{eqnarray*}%
which correspond, in these circumstances, to the naive cubic
statistics/bispectrum (without linear term), the cubic statistic with mean
subtraction, and the proper bispectrum with Wick polynomials/linear term
subtraction. Because the $X_{i}$ are exactly independent, it is trivial to
see that the CLT holds, and all three statistics are asymptotically
Gaussian. Nevertheless, it is readily seen that
\begin{eqnarray*}
Var\left\{ B_{1n}\right\}  &=&\frac{1}{n}\sum_{i=1}^{n}Var\left\{
X_{i}^{3}\right\} =\frac{1}{n}\sum_{i=1}^{n}\left\langle
X_{i}^{6}\right\rangle  \\
&=&\frac{15}{n}\sum_{i=1}^{n}\sigma _{i}^{6}\rightarrow 15\left\{ \frac{%
\sigma _{1}^{6}+\sigma _{2}^{6}}{2}\right\} =210\mbox{ ,}
\end{eqnarray*}%
because  $\left\langle X_{i}^{6}\right\rangle =15\sigma _{i}^{6}$ by Wick's
Theorem / Diagram Formula Eq.~(\ref{diagform}) (which in this case must include the so-called
flat edges). On the other hand
\begin{eqnarray*}
Var\left\{ B_{3n}\right\}  &=&\frac{1}{n}\sum_{i=1}^{n}Var\left\{
X_{i}^{3}-3\sigma _{i}^{2}X_{i}\right\}  \\
&=&\frac{1}{n}\sum_{i=1}^{n}\left\langle X_{i}^{6}-6\sigma
_{i}^{2}X_{i}^{4}+9\sigma _{i}^{4}X_{i}^{2}\right\rangle  \\
&=&\frac{1}{n}\sum_{i=1}^{n}\left\{ 15\sigma _{i}^{6}-18\sigma
_{i}^{6}+9\sigma _{i}^{6}\right\}  \\
&=&\frac{15}{n}\sum_{i=1}^{n}\sigma _{i}^{6}\rightarrow 6\left\{ \frac{%
\sigma _{1}^{6}+\sigma _{2}^{6}}{2}\right\} =84\mbox{ .}
\end{eqnarray*}%
We see thus that the linear terms is indeed not negligible ($Var\left\{
3\sigma _{i}^{2}X_{i}\right\} =9\sigma _{i}^{6},$ as compared to $Var\left\{
X_{i}^{3}\right\} =15\sigma _{i}^{6}$), and, in view of its negative
correlation with the cubic statistics, it induces a major decrease in the
variance. Concerning mean subtraction, simple computations show that%
\[
\frac{1}{\sqrt{n}}\sum_{i=1}^{n}\left( X_{i}-\overline{X}_{n}\right) ^{3}%
\mbox{ }
\]%
\[
=\frac{1}{\sqrt{n}}\sum_{i=1}^{n}X_{i}^{3}-\frac{3}{n}\frac{1}{\sqrt{n}}%
\sum_{i=1}^{n}X_{i}^{2}\left( \sum_{j=1}^{n}X_{j}\right)
\]
\[+\frac{3}{\sqrt{n}}%
\sum_{i=1}^{n}X_{i}\overline{X}_{n}^{2}-\mbox{ }\frac{1}{\sqrt{n}}%
\sum_{i=1}^{n}\overline{X}_{n}^{3}\mbox{ .  }
\]%
Now the third and fourth term are easily seen to be converge to zero, from
the law of large numbers. For the second summand, switching sums we obtain%
\[
\frac{3}{n}\frac{1}{\sqrt{n}}\sum_{i=1}^{n}X_{i}^{2}\left(
\sum_{j=1}^{n}X_{j}\right) =\frac{3}{\sqrt{n}}\sum_{i=1}^{n}X_{i}\left(
\frac{1}{n}\sum_{j=1}^{n}X_{j}^{2}\right) \mbox{ ,}
\]%
where, again by the law of large numbers we have the convergence (with
probability one)%
\[
\left( \frac{1}{n}\sum_{j=1}^{n}X_{j}^{2}\right) \rightarrow \frac{\sigma
_{1}^{2}+\sigma _{2}^{2}}{2}\mbox{ .}
\]%
Hence, neglecting terms of order $n^{-1}$ we have%
\[
B_{2n}\simeq \frac{1}{\sqrt{n}}\sum_{i=1}^{n}\left( X_{i}^{3}-3\frac{\sigma
_{1}^{2}+\sigma _{2}^{2}}{2}X_{i}\right) ^{3}\mbox{ ;}
\]%
Using the Diagram Formula of Eq.~(\ref{diagform}) again, after some manipulations one obtains%
\[
Var\left\{ B_{2n}\right\}  \rightarrow 9\sigma _{1}^{2}\times \frac{\sigma _{1}^{4}+\sigma _{2}^{4}+2\sigma
_{1}^{2}\sigma _{2}^{2}}{8}
\]
\[+9\sigma _{2}^{2}\times \frac{\sigma
_{1}^{4}+\sigma _{2}^{4}+2\sigma _{1}^{2}\sigma _{2}^{2}}{8} 
\]
\[\frac{33\sigma _{1}^{6}+33\sigma _{2}^{6}-9\sigma _{1}^{2}\sigma
_{2}^{4}-9\sigma _{2}^{2}\sigma _{1}^{4}}{8}\mbox{ ,}
\]%
so that, for $\sigma _{1}^{2}=1,$ $\sigma _{2}^{2}=3$ we have%
\[
Var\left\{ B_{2n}\right\} \rightarrow \frac{33+33\times 27-81-27}{8}=102%
\mbox{ .}
\]
The difference between the variance with mean subtraction and the linear
term is given by%
\[
Var\left\{ B_{2n}\right\} -Var\left\{ B_{3n}\right\} \rightarrow
\]%
\[
-\frac{9}{8}\sigma _{1}^{4}(\sigma _{2}^{2}-\sigma _{1}^{2})+\frac{9}{8}%
\sigma _{2}^{4}(\sigma _{2}^{2}-\sigma _{1}^{2}) \\
\]
\[=\frac{9}{8}(\sigma _{2}^{2}-\sigma _{1}^{2})^{2}(\sigma _{2}^{2}+\sigma
_{1}^{2})=18\mbox{ .}
\]

\begin{remark}
Clearly the approximation of the linear term improves when the anisotropy
decreases; in fact, the difference $Var\left\{ B_{2n}\right\} -Var\left\{
B_{3n}\right\} \rightarrow 0$ as $\sigma _{2}^{2}\rightarrow \sigma _{1}^{2}.
$ For instance, taking  $\sigma _{1}^{2}=1,$ $\sigma _{2}^{2}=2$ we have
\[
Var\left\{ B_{1n}\right\} \rightarrow 15\left\{ \frac{\sigma _{1}^{6}+\sigma
_{2}^{6}}{2}\right\} =\frac{135}{2}=67.5\mbox{ ,}
\]
\[
Var\left\{ B_{3n}\right\} \rightarrow 6\left\{ \frac{\sigma _{1}^{6}+\sigma
_{2}^{6}}{2}\right\} =27\mbox{ ,}
\]%
\[
Var\left\{ B_{2n}\right\} -Var\left\{ B_{3n}\right\} =\frac{9}{8}(\sigma
_{2}^{2}-\sigma _{1}^{2})^{2}(\sigma _{2}^{2}+\sigma _{1}^{2})=\frac{27}{8}%
\mbox{ ,}
\]
whence
\[
Var\left\{ B_{2n}\right\} \rightarrow 6\left\{ \frac{\sigma _{1}^{6}+\sigma
_{2}^{6}}{2}\right\} =30.375\mbox{ .}
\]
\end{remark}

\section{Application to $WMAP$ data}\label{application}

\subsection{The data}\label{data}

In order to test the effect of the linear term correction we applied the needlet and SMHW
estimators to the foreground reduced $V+W$ 
maps of the $WMAP$ 7-year data. For simplicity we coadded the two
frequency band maps with a constant noise weight. We analyzed the maps
at HEALPix resolution
$N_{side}=512$ and masking out galactic foregrounds and
point sources with the extended temperature analysis mask, known as
KQ75 \citep{gold11}. Where Gaussian simulations are necessary we used the 
parameters from the 
$WMAP7$+BAO+$H_0$ cosmological data to simulate the CMB sky \citep{komatsu11}, then
applying the beam and noise properties supplied by the $WMAP$ team.

\subsection{The needlets/wavelets $f_{\rm NL}$ estimator with the linear term}

The $f_{\rm NL}$ estimator based on the needlets bispectrum has been
developed and applied to $WMAP$ data in  in
\citet{rudjord09,rudjord10}. We recall the main features of this estimator.
 The needlets bispectrum can be expressed as
\beq\label{oldneedbisp}
I_{{j_1}{j_2}{j_3}}=\sum_{k}
\frac{\beta_{{j_1}{k}}\beta_{{j_2}{k}}\beta_{{j_3}{k}}}{\sigma_{{j_1}{k}}\sigma_{{j_2}{k}}\sigma_{{j_3}{k}}} ,
\eeq
where $\beta_{jk}$ is the needlet coefficient at scale $j$ and
direction, i.e. pixel, $k$,  defined in Eq.~(\ref{dirtra}), and $\sigma_{jk}$ is the expected standard deviation
of $\beta_{jk}$. The needlets bispectrum is used to estimate $f_{\rm NL}$ by a
$\chi^2$ minimization procedure:
\beq\label{chisq}
\chi^2(f_{\rm NL})={\bf d}^T(f_{\rm NL}){\bf C}^{-1}{\bf d}(f_{\rm NL})
\eeq
where the data vector is
\beq\label{datavector}
{\bf d}=I^{\rm obs}_{{j_1}{j_2}{j_3}} - f_{\rm NL} \langle \hat{I}_{{j_1}{j_2}{j_3}}  \rangle.
\eeq
Here $I^{\rm obs}$ is the bispectrum of the observed data and $\langle
\hat{I}_{{j_1}{j_2}{j_3}}  \rangle$ is the average first-order
non-Gaussian bispectrum obtained from non-Gaussian simulations. In
this analysis we used simulations with local-type non-Gaussianity
generated with the algorithm described in \citet{ngsims}. The covariance matrix $\bf C$ is obtained
by means of Monte Carlo simulations:
\beq\label{cov}
C_{ij} = \langle d_i d_j \rangle - \langle d_i \rangle \langle d_j \rangle.
\eeq
Differentiating Eq.~(\ref{chisq}) yields the estimate
\beq\label{estimator}
f_{\rm NL} = \frac{\langle \hat{I}_{{j_1}{j_2}{j_3}}  \rangle^T {\bf
    C}^{-1} I^{\rm obs}_{{j_1}{j_2}{j_3}} } { \langle \hat{I}_{{j_1}{j_2}{j_3}}  \rangle^T {\bf
    C}^{-1} \langle \hat{I}_{{j_1}{j_2}{j_3}}  \rangle}
\eeq 
Details of the estimation procedure can be found in \citet{rudjord09}.

We want now to adopt the linear term correction in order to decrease
the variance of the estimator (\ref{estimator}). Following Eq.~({\ref{need_linterm}), it is straightforward to subtract the linear term
from the bispectrum in Eq.~(\ref{oldneedbisp}). The needlet bispectrum
$I^{\rm obs}_{{j_1}{j_2}{j_3}}$ in the data vector (\ref{datavector}) can be expressed now as

\beqar
&&I^{\rm obs}_{{j_1}{j_2}{j_3}}= 
\frac{1}{\sigma _{j_{1}}\sigma _{j_{2}}\sigma _{j_{3}}}
\sum_{k} \{ \beta _{j_{1}k}\beta _{j_{2}k}\beta _{j_{3}k}
\nonumber\\ 
&-&\Gamma _{j_{1}j_{2}}(k)\beta _{j_{3}k} -\Gamma
_{j_{1}j_{3}}(k)\beta _{j_{2}k} -\Gamma
_{j_{2}j_{3}}(k)\beta _{j_{1}k_{1}}\},\label{newneedbisp}
\eeqar
where
\beq
\Gamma _{j_{1}j_{2}}(k)=\left\langle \beta _{j_{1}k}\beta
_{j_{2}k}\right\rangle \text{ ,}
\eeq
and similarly for the SMHW where the $\beta_{jk}$ are replaced by the $w(R,n_k)$.
We note that the contribution of the linear term to the average
$\langle\hat{I}_{{j_1}{j_2}{j_3}} \rangle$ in the data vector
(\ref{datavector}) is vanishing, as the
expected linear term value is always zero (Eq.~(\ref{expvalue_linterm})).
The terms $\Gamma _{j_{1}j_{2}}(k)$ contains one contribution from the CMB and one from the noise. 
The former is calculated using Monte Carlo simulations, the latter is calculated analytically.\\
We implemented slightly different algorithms of the $f_{\rm NL}$ estimator
described above. The cubic statistic in Eqs.~(\ref{oldneedbisp}, \ref{newneedbisp}) has been obtained in three cases, 
namely with standard needlets, Mexican needlets and SMHW coefficients.

\subsection{Analysis of the $WMAP$ data}

In order to test the effect of the linear term correction, we applied the $f_{\rm NL}$ 
estimator to the $WMAP$ data described in Section~\ref{data}, both before
and after the addition of the linear term. The standard needlets
coefficients has been obtained with two different bases: $B=1.781$
with scales $j=1-13$ and $B=1.34$ with $j=3-22$. For the Mexican
needlets we chose $B=1.34$ and $p=1$ with $j=3-22$. In all the cases in addition
to the needlet scales we considered a ``scale0'', i.e. the original
map not convolved with the needlets.  The scales selected for the SMHW
are the same scales used in \citet{curto2,curto3}: $R_0=0$ (the
uncolvolved map), $R_1=2.9, R_2=4.5, R_3=6.9, R_4=10.6, R_5=16.3,
R_6=24.9, R_7=38.3, R_8=58.7, R_9=90.1, R_{10}=138.3, R_{11}=212.3,
R_{12}=325.8, R_{13}=500., R_{14}=767.3$ arc minutes. The SMHW
coefficients have been analyzed in two ways: as they are or after subtracting
the scale-by-scale average of the coefficients outside the
applied mask. Thereafter we tested the same ``mean subtraction'' procedure with standard and Mexican needlets. All the analysis are performed up to the multipole $\ell=1500$.\\
\begin{table}[!h]
\centering
\caption{Masked fraction of the sky with the extended KQ75 masks} \label{mask}
\begin{tabular}{cc|cc}
\hline\hline
\multicolumn{2}{c}{SMHW}&\multicolumn{2}{c}{Mexican needlets}\\
scale $R_i$& \% & scale $j$ & \%\\
\hline
0 (map) & 29.4 & 1 & 66.2 \\
1  &  29.4 & 2& 60.9 \\
2  &    29.4& 3&55.8 \\
3   &   29.4& 4&  51.0 \\ 
4   &   29.4& 5& 45.9 \\ 
5    &  29.4& 6& 41.8 \\ 
6   &   29.4& 7& 39.1 \\ 
7  &    29.6& 8& 36.9 \\
8   &   30.6& 9& 35.3 \\
9   &   33.3& 10& 34.2 \\
10  &    37.8& 11& 33.4 \\
11  &    44.5& 12& 32.7  \\
12  &    54.0& 13& 32.3 \\
13   &   67.4& 14& 32.0 \\
14   &   83.4& 15& 31.6 \\
&& 16& 31.3 \\
&& 17& 31.0 \\
&& 18& 30.7 \\
&& 19&30.4  \\
&& 20& 30.2 \\
&& 21& 30.1 \\
&&22 & 30.0\\
&& scale0& 29.4\\
\hline\hline
\end{tabular}
\end{table}
With the exception of one case -- where we analyzed the
full-sky with standard needlets $B=1.781$ -- we always applied the KQ75
mask. For the SMHW analysis it is necessary to properly extend the
mask at each scale, because the pixels near the border are affected by
the zero values of the cut. Given the similarity with the
SMHW, for comparison we extended the mask also for one case with the Mexican needlets. Details of
the mask extension procedure can be found in \citet{mcewen05}. The fractions of masked
sky obtained with the extended masks at each scale are showed in Table~\ref{mask}. \\
For each analyzed case we simulated a set of 51000 Gaussian
maps with the same beam and noise properties of the $WMAP$ coadded $V+W$ cleaned map. 
40800 simulations have been used to obtain the covariance matrix in
Eq.~(\ref{cov}).  The standard deviation of the remaining 10200 simulations gives the error bar associated with the estimated $f_{\rm  NL}$  values. 
Moreover, in order to verify that the $f_{\rm NL}$ estimator is unbiased, for
each case we also analyzed a set of 700 non-Gaussian simulations with
an input $f_{\rm NL}=30$ \citep{ngsims}. All the results are reported
in Table ~\ref{results}.\\
\begin{table}[!h]
\centering
\caption{Results}\label{results}
\begin{tabular}{c|ccccc}
\hline\hline
\\
case&linear &  $WMAP$ & Error  & $\Delta\sigma$ &NG sims.\\
& term & $f_{\rm NL}\,^a$ & Bar $1\sigma^b$& \%$^c$& $\langle f_{\rm NL}\rangle\,^d$ \\
\\
\hline
\multicolumn{6}{c}{\,}\\
\multicolumn{6}{c}{standard needlets:}\\
\hline 
$B=1.781$ & no & // & 18.4 & &29.2 \\
   fullsky  & yes & // & 18.3 & 0.5 &29.2 \\
\hline
$B=1.781$   & no & 63.5 & 25.4 & &29.9 \\
    KQ75 & yes& 41.5 & 22.2 & 12.6 &30.0 \\
\hline
$B=1.34$     & no & 43.6 & 24.9 & &29.3 \\
      KQ75      & yes & 39.3 & 22.1 & 11.2 &29.5 \\
\multicolumn{6}{c}{\,}\\
\multicolumn{6}{c}{Mexican needlets:}\\
\hline
$B=1.34, p=1$      & no & 37.8 & 25.4 & &29.5 \\
   ext.   KQ75    & yes& 26.6 & 22.2 & 12.6 &30.3 \\
\hline 
$B=1.34, p=1$      & no & 39.2 & 23.2 & &29.9\\
     KQ75    & yes& 37.5 & 21.8 & 6.0 &29.8 \\
\multicolumn{6}{c}{\,}\\
\multicolumn{6}{c}{SMHW:}\\
\hline 
no mean subtr. & no & 77.8 & 27.3 & &29.7 \\
     ext. KQ75            & yes & 33.1 & 21.9 & 20.1 & 30.1 \\
\hline
mean subtraction & no & 37.5 & 22.3 & 18.3$^e$&29.5 \\
    ext. KQ75         & yes & 34.4 & 22.0 & 1.3 &29.7 \\
\multicolumn{6}{c}{\,}\\
 \multicolumn{6}{c}{needlets - mean subtraction:}\\
 \hline
 standard&&&&&\\
 $B=1.34$     & no &  33.0 & 22.5 & $9.6^e$  & 29.5\\
      KQ75      & yes &  37.1 & 22.3 & 0.8 & 29.5\\
\hline
Mexican&&&&&\\
$B=1.34, p=1$      & no & 33.2 & 22.1 & $4.7^e$ & 30.0\\
     KQ75    & yes& 37.6 & 22.0 & 0.4 & 29.8\\
\hline
\hline
\end{tabular}\\
$^{(a)}$ local $f_{\rm NL}$ on coadded $V+W$ foreground reduced maps;\\
$^{(b)}$ standard deviation over 10200 Gaussian simulations;\\
$^{(c)}$ [$\sigma$(no linear term) - $\sigma$(with linear term)]/$\sigma$(no linear term);\\
$^{(d)}$ average over 700 simulations with input $f_{\rm NL}=30$;\\
$^{(e)}$ comparison with the ``no mean subtracted" $\sigma$.\\
See the text for further details.
\end{table}

\subsection{Results}
The addition of the linear term achieves a decrease in the standard deviation in all the cases. The full-sky 
analysis shows only a little improvement of the error bar - $0.5\%$ - indicating that the mask gives the
 major contribution to the correction operated by the linear term. Applying the KQ75 mask, all the 
 $f_{\rm NL}$ values estimated on the $V+W$ foreground reduced maps are consistent with the 
 $WMAP$ 7-year best estimate of $f_{\rm NL}=32 \pm 21$ \citep{komatsu11}.  We began the
analysis choosing the same needlet base $B=1.781$ as in \citet{rudjord10}. With respect to the previous
result - $f_{\rm NL}=73 \pm 31$ - we already obtained a $18\%$ smaller error bar analyzing the 7-year 
data in place of the 5-year release and using more scales. But a further improvement of $12.6\%$ is 
given by the linear term, demonstrating that the correction is non-negligible. 
The average $\langle f_{\rm NL}\rangle$ over the 700 non-Gaussian 
simulations shows that the estimator is unbiased. We can note also that the $f_{\rm NL}$ value 
estimated with the correction is closer to the $WMAP$  result than without the linear term. We then chose 
a smaller needlet base $B=1.34$, and therefore we constructed the cubic statistic of 
Eqs.~(\ref{oldneedbisp}, \ref{newneedbisp}) with more needlet scales.  We indeed note a smaller 
standard deviation of $24.9$ even without the linear correction with respect to $B=1.781$. The error bar is decreased to $22.1$, i.e. 
another $11.2\%$, adding the linear term. 

Moving to the Mexican needlets case, we started with a conservative analysis applying an extended 
KQ75 mask as for the SMHW case. Despite the reduced sky coverage (Table \ref{mask}) we found the 
same $1\sigma$ error bars obtained with standard needlets - $B=1.781$ (but $f_{\rm NL}$ values closer 
to $WMAP$).  Motivated by the results of \citet{sandro} that has shown negligible neighbour bias effects 
of the mask, we repeated the Mexican needlets analysis with the original KQ75 mask. Looking at the 
results on the non-Gaussian simulations, we notice that the $\langle f_{\rm NL}\rangle$ are indeed 
unbiased. In this case the linear term correction leads to a $6\%$ improvement of the error bar, sufficient 
to achieve our best estimate $f_{\rm NL}=37.5 \pm 21.8$. This standard deviation is very close to the $WMAP$ result 
with the optimal KSW estimator, where $\sigma=21$. With the addition of the linear term correction the 
needlets  $f_{\rm NL}$ estimator is almost optimal. We believe that further standard deviation reductions 
can be achieved with an optimal $V+W$ coadding and with a different choice of the needlet base $B$ 
and scales $j$. Anyway these exploitations are behind the primary scope of this work.\\
Furthermore we consider the case with the SMHW coefficients. 
The extensions of the KQ75 mask returns sky coverage 
percentages close to \citet{curto2,curto3}. Without subtracting the scale-by-scale 
coefficients average, the linear term correction leads to a reduction of the error bar from $27.3$ to $21.9$, equal to $20.1\%$. 
With a very similar SMHW analysis on coadded $V+W$ 7-yr cleaned maps, \citet{curto3} found 
$f_{\rm NL}=32.5 \pm 23$ (Fisher matrix bound $\sigma_F=22.5$), where here we considered the result 
uncorrected from point-source contribution. Therefore the linear term addition rather improves their 
result. However the analysis of \citet{curto3} is performed after the scale-by-scale mean subtraction.
Following the same procedure, we indeed achieved a reduction of the error bar without the linear 
term correction: from $27.3$ to $22.3$, equal to $18.3\%$. The $\sigma$ found subtracting the 
mean is indeed closer to the one of \citet{curto3}. Anyway we note that also in this case the addition 
of the linear term can still slightly improve the error bar ($1.3\%$), and the $\sigma$ found with this correction - subtracting or not subtracting the mean -  
is almost identical. \\
Motivated by the SMHW analysis, we tested the scale-by-scale mean subtraction procedure 
with both the standard and the Mexican needlets. We found that also in the needlets case this procedure
well approximates the linear term correction, providing an error bar reduction of $9.6\%$ (i.e. from $24.9$ to $22.5$) and $4.7\%$ (i.e. from $23.2$ to $22.1$) for 
standard and Mexican needlets respectively. But again the addition of linear term provides even smaller 
error bars: $0.8\%$ and $0.4\%$ respectively. Moreover we obtained the smallest $\sigma$ with the linear term but without subtracting the mean. We indeed got $22.1$ 
against $22.3$ for the standard needlets, and $21.8$ against $22.0$ for the Mexican needlets.

\section{Conclusions}\label{conclus}
In this paper, we have derived for the first time the linear correction term
for wavelets/needlet $f_{\rm NL}$ estimators. As expected, under
ideal experimental circumstances (isotropic noise, no masked regions)
this term turns out to be identically zero. Under a realistic experimental
set-up, the term is non-negligible, although smaller than for the KSW
estimators; this is due to the localization properties of wavelets/needlets
statistics, which soften to some extent the effects of unobserved regions
and anisotropic noise. We have also argued that the linear correction term
is well-approximated by scale-by-scale mean subtraction, thus providing an
explanations for recent results from \citep{curto2, curto4}, where 
numerical estimates on the variance of wavelet estimators where shown to be very close to the KSW bound; 
in fact, we have shown that mean subtraction can cover approximately 85-90\% of the linear term effect under
$WMAP$-like circumstances. The procedures we advocate are applied to
$WMAP$ 7-year $V+W$ foreground cleaned
data, confirming that the corrections achieved are non-negligible.
Applying the linear term correction we obtained the 
best estimate $f_{\rm NL}=37.5 \pm 21.8$. The error bar is very close to the optimal
standard deviation $\sigma=21$ found by the $WMAP$ team \citep{komatsu11}.\\
In view of these results, we argue that wavelets/needlets statistics can
provide a statistically sound and computationally convenient technique for
non-Gaussianity analysis on CMB data. 

\acknowledgments
FKH acknowledges an OYI grant from the Norwegian Research Council; research by DM is supported by the European Research Council grant n.277742 (\emph{Pascal}). 
This research has been partially supported by the ASI/INAF Agreement I/072/09/0 for the Planck LFI Activity of Phase E2 and by the PRIN 2009 project "La Ricerca di non-Gaussianit\`a Primordiale". 
Super computers from NOTUR (The Norwegian metacenter for computational science) have been used in this work. We acknowledge the use of the HEALPix software package \citep{healpix} 
and the Legacy Archive for Microwave Background Data Analysis (LAMBDA) to retrieve the $WMAP$ data set.

\end{document}